\documentclass[prd,aps,oamsmath,amssymb,nofootinbib,preprintnumbers]
{revtex4}
\usepackage{graphicx}
\usepackage{dcolumn}
\usepackage{bm}
\usepackage{amsmath}
\usepackage{amsthm}
\usepackage{multirow}

\usepackage{amsfonts}
\usepackage{euscript,bbm}
\usepackage{ifthen}
\usepackage{psfrag}
\usepackage{slashed}
\usepackage{hyperref}
\usepackage{gensymb}

\def\ls{\mathrel{\lower4pt\vbox{\lineskip=0pt\baselineskip=0pt
           \hbox{$<$}\hbox{$\sim$}}}}
\def\gs{\mathrel{\lower4pt\vbox{\lineskip=0pt\baselineskip=0pt
           \hbox{$>$}\hbox{$\sim$}}}}
\def\drawbox#1#2{\hrule height#2pt

\hbox{\vrule width#2pt height#1pt \kern#1pt
              \vrule width#2pt}
              \hrule height#2pt}

\def\Asym#1#2{\vcenter{\vbox{\drawbox{#1}{#2}
              \kern-#2pt       
              \drawbox{#1}{#2}}}}

\newcommand{\be}{\begin{equation}}
\newcommand{\ee}{\end{equation}}
\newcommand{\bea}{\begin{eqnarray}}
\newcommand{\eea}{\end{eqnarray}}
\providecommand{\e}[1]{\ensuremath{\times 10^{#1}}}

\newcommand{\gsim}{\lower.7ex\hbox{$\;\stackrel{\textstyle>}{\sim}\;$}}
\newcommand{\lsim}{\lower.7ex\hbox{$\;\stackrel{\textstyle<}{\sim}\;$}}

\newcommand{ \pythia } {{\tt PYTHIA 8.201}}

\newcommand{ \PPPC } {{\tt PPPC4DMID}}

\newcommand{\braket}[1]{\left<#1\right>}

\begin{document}

\begin{flushright}
MI-TH-1510
\end{flushright}

\title{Confronting Galactic center and dwarf spheroidal gamma-ray observations with cascade annihilation models}

\author{Bhaskar Dutta}
\author{Yu Gao}
\author{Tathagata Ghosh}
\author{Louis E. Strigari}

\affiliation{Mitchell Institute for Fundamental Physics and Astronomy, \\
Department of Physics and Astronomy, Texas A\&M University, College Station, TX 77843-4242, USA
}

\begin{abstract}
Many particle dark matter models predict that the dark matter undergoes cascade annihilations, i.e. the annihilation products are 4-body final states. In the context of model-independent cascade annihilation models, we study the compatibility of the dark matter interpretation of the Fermi-LAT Galactic center gamma-ray emission with null detections from dwarf spheroidal galaxies. For canonical values of the Milky Way density profile and the local dark matter density, we find that the dark matter interpretation to the Galactic center emission is strongly constrained. However, uncertainties in the dark matter distribution weaken the constraints and leave open dark matter interpretations over a wide range of mass scales. 
\end{abstract}

\maketitle


\section{Introduction} 



Analyses of Fermi-LAT data by several groups have identified an emission of diffuse gamma rays distributed nearly spherically-symmetric about the Galactic center, i.e. the Galactic  Center Excess (GCE)~\cite{Hooperon,GCE-others,Daylan:2014rsa},~\cite{Calore:2014xka} (CCW). The GCE is statistically significant, though its precise morphology and energy spectrum is still subject to systematic uncertainties that derive from the model fits. Millisecond pulsars~\cite{Pulsar}, young pulsars~\cite{O'Leary:2015gfa}, and more generally a population of point source below the Fermi-LAT threshold~\cite{Bartels:2015aea,Lee:2015fea} have been fit to the GCE. Other astrophysical sources such as cosmic ray protons~\cite{Carlson:2014cwa} and inverse Compton emission from high energy electrons~\cite{Petrovic:2014uda,Cholis:2015dea,Gaggero:2015nsa} arising from  burst-like events have also been discussed in the context of the GCE.

A dark matter (DM) annihilation explanation of the GCE has generated considerable excitement~\cite{Calore:2014nla, Anti-pulsar, Agrawal:2014oha, Cerdeno:2015ega, Hooper-4body, 4body, Cline:2015qha,GCE-DM}. In itself, there are a couple of challenges one must confront when attempting to connect the GCE to a possible DM signal. First, the aforementioned emission from unresolved point sources and diffuse emission process are difficult to predict theoretically, which implies that the data itself is often used to understand the gamma ray emission from these sources. Second, there is considerable freedom in DM interpretations of the excess, in that a wide range of masses and cross sections are able to fit the data. 

With these points in mind, studies of other sources for a corroborating DM signal are especially important. 
Dwarf spheroidal galaxies (dSphs) of the Milky Way are quintessential target for indirect DM searches~\cite{Strigari:2013iaa,Conrad:2015bsa}, and provide an independent cross check on a possible DM signal hinted at near the Galactic center. Indeed the lack of excess gamma-ray signal from dSphs imposes constraints on DM annihilation cross-section~\cite{Ackermann:2015zua}, and also strongly constrains DM interpretations of the GCE for a variety of different annihilation channels with 2 body final states. 

In this paper we explore DM particle models that annihilate to a pair of on-shell scalar mediators which subsequently decay into $b$-quarks and $\tau$ leptons, and explore their compatibility with GCE and dSph gamma ray observations. Annihilation to 4-body final states have been considered within the context of earlier Fermi-LAT dSph observations and earlier analyses of the GCE~\cite{Hooper-4body, 4body}. In comparison to these previous papers, the goal of the paper is two-fold. First we revisit the annihilation of DM into higgs-like scalars, taking into account correlated systematic uncertainties derived by CCW.  We then constrain the model parameter space using the new Fermi-LAT dSph {\tt Pass-8} results~\cite{Ackermann:2015zua}. A similar study, prior to recent {\tt Pass-8} results, has been performed in the context of the NMSSM~\cite{Cerdeno:2015ega}. In contrast, in this paper we fit the GCE in $4b$, $4 \tau$ and $2b \, 2 \tau$ channels in both a model-independent way and within the framework of a  realistic  $U(1)_{B-L}$ model incorporating all of the aforementioned decay channels.
 
The structure of the paper is as follows. In Section~\ref{setup} we discuss our fits to the Galactic center emission and the framework for the statistical analysis. In Section~\ref{dSphs}, we interpret the new dSph constraints in the context of our analysis. In Section~\ref{Results} we present the results of our model-independent study. In Section~\ref{Model} we describe the motivation and particle content of the $U(1)_{B-L}$ model along with its possible realization in the context of GCE phenomenology. Finally we conclude in Section~\ref{Conclusion}.

\section{\label{setup}Fitting the GCE with cascade annihilation through a scalar}

The direct production of hard photons from DM annihilation is typically loop suppressed~\cite{Bergstrom:2012fi}, so that photons produced are from decays of Standard Model (SM) particles. Here we consider the DM particle, $\chi$, annihilating to a pair of beyond Standard Model (BSM) scalar, $\phi$, which in turn decays to various SM quarks and leptons. The continuous spectrum of gamma-rays arises from light mesons, produced via hadronization and/or decay of SM fermions.

The gamma-ray differential flux from DM annihilation over a solid angle $\Delta \Omega$ is given by,
\begin{align}
\dfrac{d \Phi^{\gamma}}{dE_{\gamma}} = \dfrac{1}{4 \pi }\dfrac{\braket{\sigma v}}{m^2_{\chi}}  \sum\limits_{f}  \dfrac{d N^{\gamma}_f}{dE_{\gamma}}  \times \dfrac{1}{\Delta \Omega}\int_{\Delta\Omega} \int_{l.o.s} \rho^2(r(s,\psi)) \,ds \,d\Omega \,, 
\end{align}
where the sum is extended over all annihilation channels into fermionic final states $f$. The first term depends on particle physics properties - $\braket{\sigma v}$ is the thermally averaged total cross section, $m_{\chi}$ is the DM mass, and $\dfrac{d N^{\gamma}_f}{dE_{\gamma}}$ is the prompt photon spectrum per annihilation into final state $f$. The second term, known as the astrophysical J-factor, is obtained from the line of sight (l.o.s) integration over DM halo profile, $\rho(r(s,\psi))$, where $r(s,\psi) = \sqrt{r^2_{\odot}+s^2-2 r_{\odot} s \cos{\psi}}$, with $r_{\odot} = 8.5$ kpc and $\psi$ being the angle from the galactic center. To provide the most straightforward comparison to previous results we utilize the generalized Navarro-Frenk-White (gNFW) profile for the DM distribution~\cite{gNFW}
\begin{align}
\rho(r) = \rho_0 \dfrac{(r/r_s)^{-\gamma}}{(1+r/r_s)^{3-\gamma}} \, ,
\end{align} 
The scale radius $r_s$, is chosen to be 20 kpc, and the scale density $\rho_0$ is determined by fixing the local DM density at the Solar radius, $\rho_\odot = 0.4$ GeV/cm$^3$~\cite{Bovy:2012tw}. For a DM interpretation of the GCE, the best fit is $\gamma = 1.2$~\cite{Daylan:2014rsa,Calore:2014xka} over a region of interest (ROI) $2 \degree \leq |b| \leq 20 \degree$ and $|l| \leq 20 \degree$. For these assumptions the averaged J-factor over the ROI, $\bar{J}$, is found to be $2.06\e{23}$ GeV$^2$cm$^{-5}$sr$^{-1}$.

To fit to the GCE, we use the results of CCW, who go into detail exploring multiple Galactic diffuse emission (GDE) models. The aforementioned analysis has been implemented by generating the prompt photon spectra using \pythia~\cite{Sjostrand:2014zea} and we verified that our results agree with \PPPC~\cite{PPPC} for $b\bar{b}$ and $\tau^+ \tau^-$ final states. Next we have performed a global fit using a $\chi^2$ statistic defined by,
\begin{align}
\chi^2 = \sum\limits_{ij}(\dfrac{d\Phi^{\gamma}_i}{dE_{\gamma}} - \dfrac{dF_i}{dE_{\gamma}}) (\Sigma^{-1})_{ij} (\dfrac{d\Phi^{\gamma}_j}{dE_{\gamma}} - \dfrac{dF_j}{dE_{\gamma}}) \, ,
\end{align}
where $\dfrac{d\Phi^{\gamma}_i}{dE_{\gamma}}$ and $\dfrac{dF_i}{dE_{\gamma}}$ are the predicted and observed flux in the $i$-th energy bin and $\Sigma_{ij}$ is the covariance matrix containing statistical and correlated systematic errors. CCW have estimated the uncertainties of the GCE by studying 60 GDE models and also studied the correlation in the spectrum along the Galactic disc. CCW extract the residual signal and a set of systematic uncertainties, which dominates over the energy range of our interest and has high degree of correlation across energy bins. The effect of these systematic uncertainties are included in our analysis by means of the publicly available covariance matrix, $\Sigma_{ij}$~\cite{CCW}. 

\section{\label{dSphs}Extracting constraints from dwarf spheroidal galaxies}

From a combined sample of 15 dSphs, Fermi-LAT has presented the upper bounds on $\braket{\sigma v}$ in standard SM annihilation channels [$l^+ l^-(l= e,\mu,\tau), \, u\bar{u}, \, b\bar{b}, \, W^+W^-$] based on six years of data~\cite{Ackermann:2015zua}. These results have improved the cross section constraints derived from previous combined samples~\cite{GeringerSameth:2011iw,Ackermann:2011wa,Ackermann:2013yva,Geringer-Sameth:2014qqa}. Our goal is to use these bounds to estimate the sensitivity to 4-fermion final states. 

\begin{figure}[!t]
\centering
\includegraphics[scale=.2]{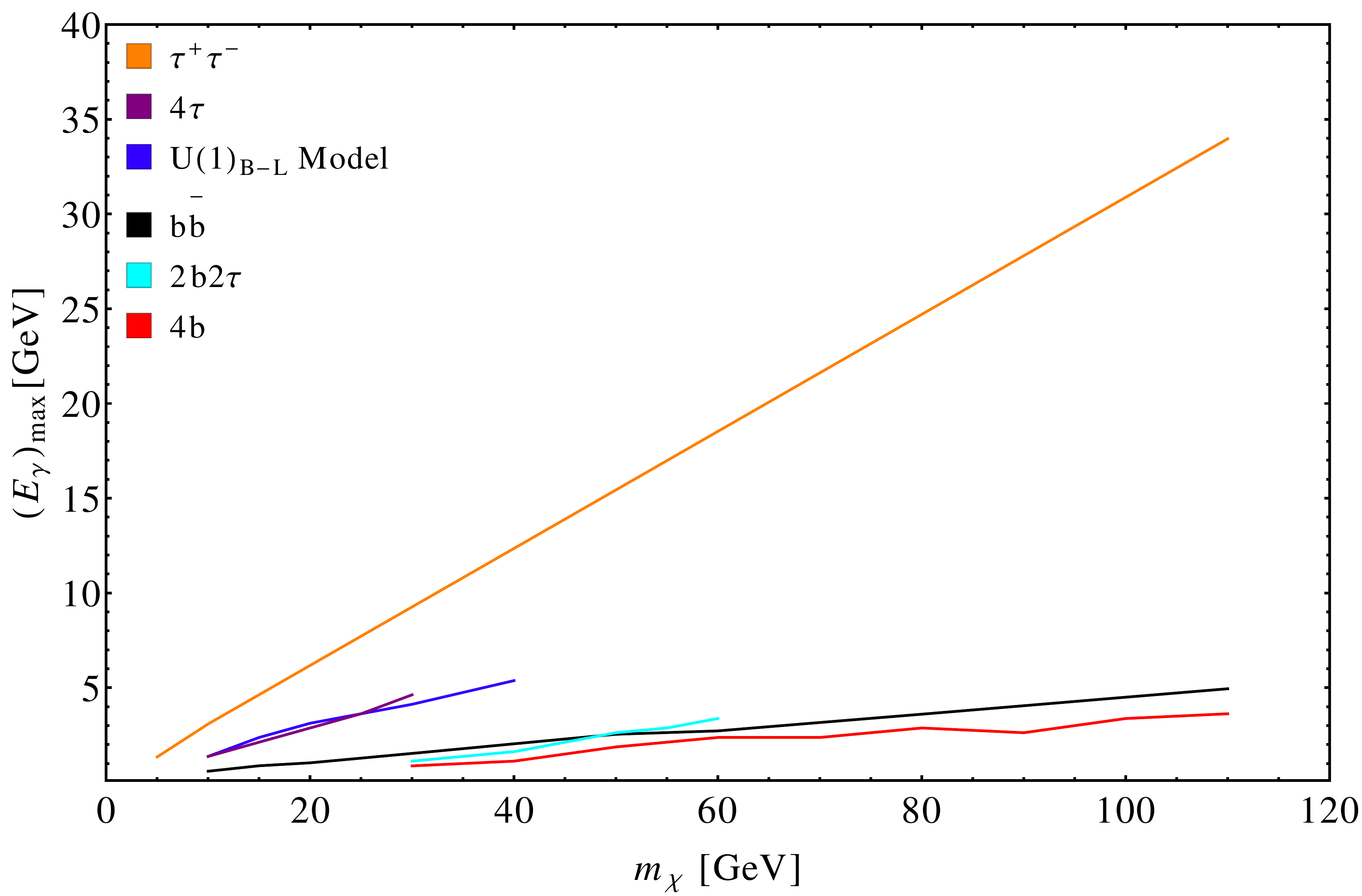}
\caption{The peak of $E^2_{\gamma} \, \dfrac{dN^{\gamma}}{dE_{\gamma}}$, which we define as $(E_{\gamma})_{max}$, as a function of DM mass, $m_{\chi}$, for all 2-body and 4-body channels. The $U(1)_{B-L}$ model is described in Section~\ref{Model}. }
\label{Emax}
\end{figure}

To deduce constraints on 4-body final states from the published Fermi-LAT constraints on 2-body final states we utilize the following procedure. For each of the 2-body and 4-body final states that we consider we calculate the photon spectrum, $\dfrac{dN^{\gamma}}{dE_{\gamma}}$, and for each spectrum identify the peak energy of $E_\gamma^2 \dfrac{dN^{\gamma}}{dE_{\gamma}}$, which we define as $(E_{\gamma})_{max}$. Our motivation for this definition of $(E_{\gamma})_{max}$ comes from the fact that different channels with the same $(E_{\gamma})_{max}$ have roughly the same spectral shape. For all 2-body and 4-body channels, in Figure~\ref{Emax} we show $(E_{\gamma})_{max}$ as a function of the DM mass.

\begin{figure}[!t]
\centering
\includegraphics[scale=.2]{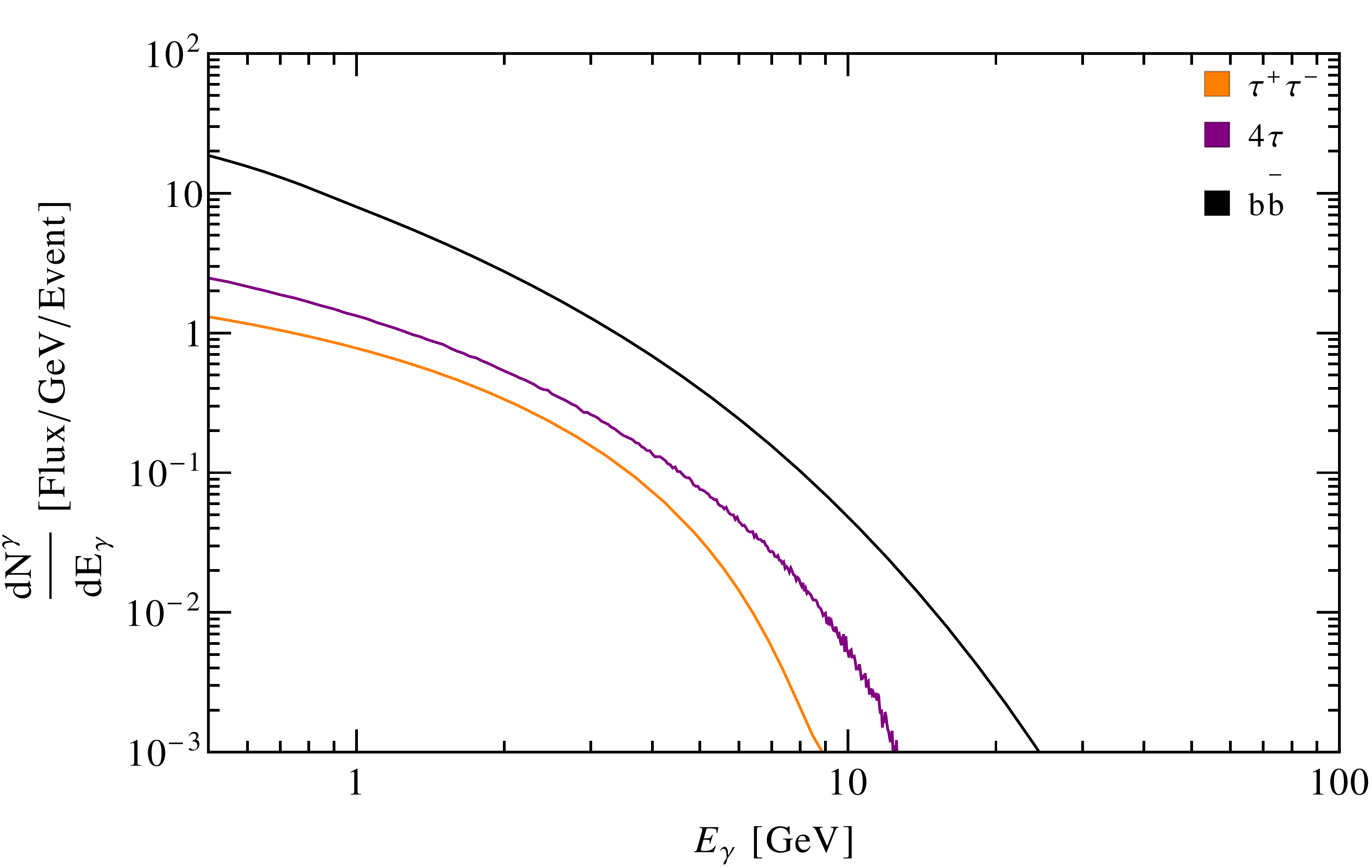}
\includegraphics[scale=.2]{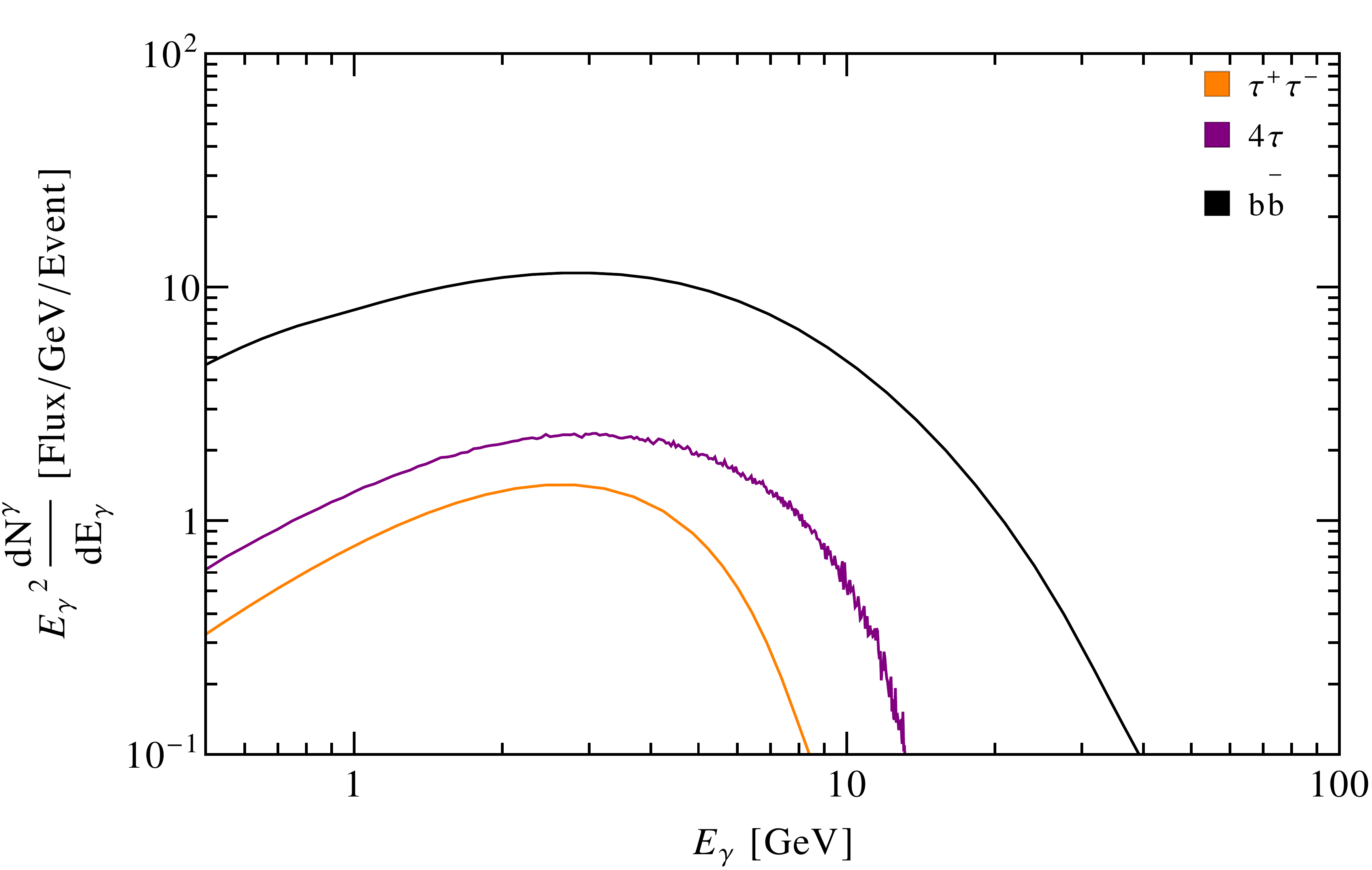}
\caption{The photon energy spectrum, $\dfrac{dN^{\gamma}}{dE_{\gamma}}$, and the spectrum weighted by the energy-squared, $E^2_{\gamma} \, \dfrac{dN^{\gamma}}{dE_{\gamma}}$, for $4 \tau$ (purple), $\tau^+\tau^-$ (orange) and $b \bar{b}$ (black) final states that have the same $(E_{\gamma})_{max}$. For $4 \tau$ channels the DM mass is $m_{\chi} = 19$ GeV, for the $\tau^+\tau^-$ channel it is $m_{\chi} = 9$ GeV, and for the $b \bar{b}$ channel it is $m_{\chi} = 59$ GeV. 
}
\label{Shape}
\end{figure}

As an example in Fig.~\ref{Shape} we show the shape of both $\dfrac{dN^{\gamma}}{dE_{\gamma}}$ and $E^2_{\gamma} \, \dfrac{dN^{\gamma}}{dE_{\gamma}}$ for the $4 \tau$ final state with $m_{\chi} = 19$ GeV, for the $\tau^+ \tau^-$ final state with $m_{\chi}=9$ GeV, and for the $b \bar{b}$ final state with $m_{\chi}=59$ GeV. All of these final states have the same $(E_{\gamma})_{max}$. These figures show that at asymptotically low and high photon energies, the shapes of both $\dfrac{dN^{\gamma}}{dE_{\gamma}}$ and $E^2_{\gamma} \, \dfrac{dN^{\gamma}}{dE_{\gamma}}$ for the spectra with similar $(E_{\gamma})_{max}$ are similar. 

More generally, we derive an upper bound on $\braket{\sigma v}$ at a given mass $m_{\chi}^{4-body}$ for a 4-body final state by matching its $(E_{\gamma})_{max}$ with the corresponding $(E_{\gamma})_{max}$ of a SM 2-body final state, at a mass $m_{\chi}^{2-body}$. We determine the ratio of the total flux in our 4-body final state to the total flux from the 2-body SM final state, 
\begin{align}
\braket{\sigma v}_{4-body} = \braket{\sigma v}_{2-body} \times \bigg(\dfrac{m_{\chi}^{4-body}}{m_{\chi}^{2-body}}\bigg)^2 \times \int^{m_{\chi}^{2-body}}_{0.5 \, GeV} \frac{d \Phi^{\gamma}_{2-body}}{dE_{\gamma}}dE_{\gamma} \times \bigg( \int^{m_{\chi}^{4-body}}_{0.5 \, GeV} \frac{d \Phi^{\gamma}_{4-body}}{dE_{\gamma}}dE_{\gamma} \bigg)^{-1} \, .
\label{eq:scaling} 
\end{align}
The lower photon energy limit of 0.5 GeV is motivated by the lower energy cut-off in the Fermi-LAT dSphs study~\cite{Ackermann:2015zua}.

The simple approach we have outlined above is used to extract plausible bounds on 4-body states without having to run through a full maximum likelihood analysis. One question that we must address is how our 4-body final state upper bounds on $\braket{\sigma v}$ depend on the particular choice of 2-body final state that we use for the scaling in Equation~\ref{eq:scaling}. To answer this question we have tested all 2-body scaling channels, and we generally find that the bounds obtained scaling to the $\tau^+ \tau^-$ and 
$b \bar{b}$ 2-body final states agree within $20 \%$ of each other. In the next section our bounds are discussed in detail. 

\newpage

\section{\label{Results}Results}
We consider the scenario in which DM particles annihilate to produce a pair of scalars, $\phi$, which then decay into a pair of quarks and leptons. For the decay of $\phi$, we first explore three model independent scenarios,  
\begin{eqnarray}
\chi \chi &\rightarrow& \phi \phi,  \, (\phi \rightarrow b \bar{b}) \\ 
\chi \chi &\rightarrow& \phi \phi,  \, (\phi \rightarrow \tau^+ \tau^-)\\
\chi \chi &\rightarrow& \phi_1 \phi_2, \, (\phi_1 \rightarrow b \bar{b}, \, \phi_2 \rightarrow \tau^+ \tau^-). \label{eq:iii} 
\end{eqnarray}
For simplicity $\phi_1$ and $\phi_2$ are assumed to be degenerate in mass in the case of Equation~\ref{eq:iii}. As a working example of these scenarios in Section~\ref{Model} we discuss a $U(1)_{B-L}$  model, in which $\phi$ decays to $b\bar{b}$ and $\tau^+ \tau^-$ channels with different branching ratios (BR), depending on the $\phi$ mass. To compare these 4-body final states to more standard 2-body final state models, we have fit the GCE in canonical $b \bar{b}$ and $\tau^+ \tau^-$ annihilation channels, and for 2-body channels we find good agreement with previous results~\cite{Calore:2014nla,Agrawal:2014oha,Cerdeno:2015ega}. Moreover we compare our results for $4b$ and $4 \tau$ final states to previous studies~\cite{Cerdeno:2015ega}, and find good agreement in regions where the parameter spaces coincide. 

\begin{figure}[!t]
\centering
\includegraphics[scale=.3]{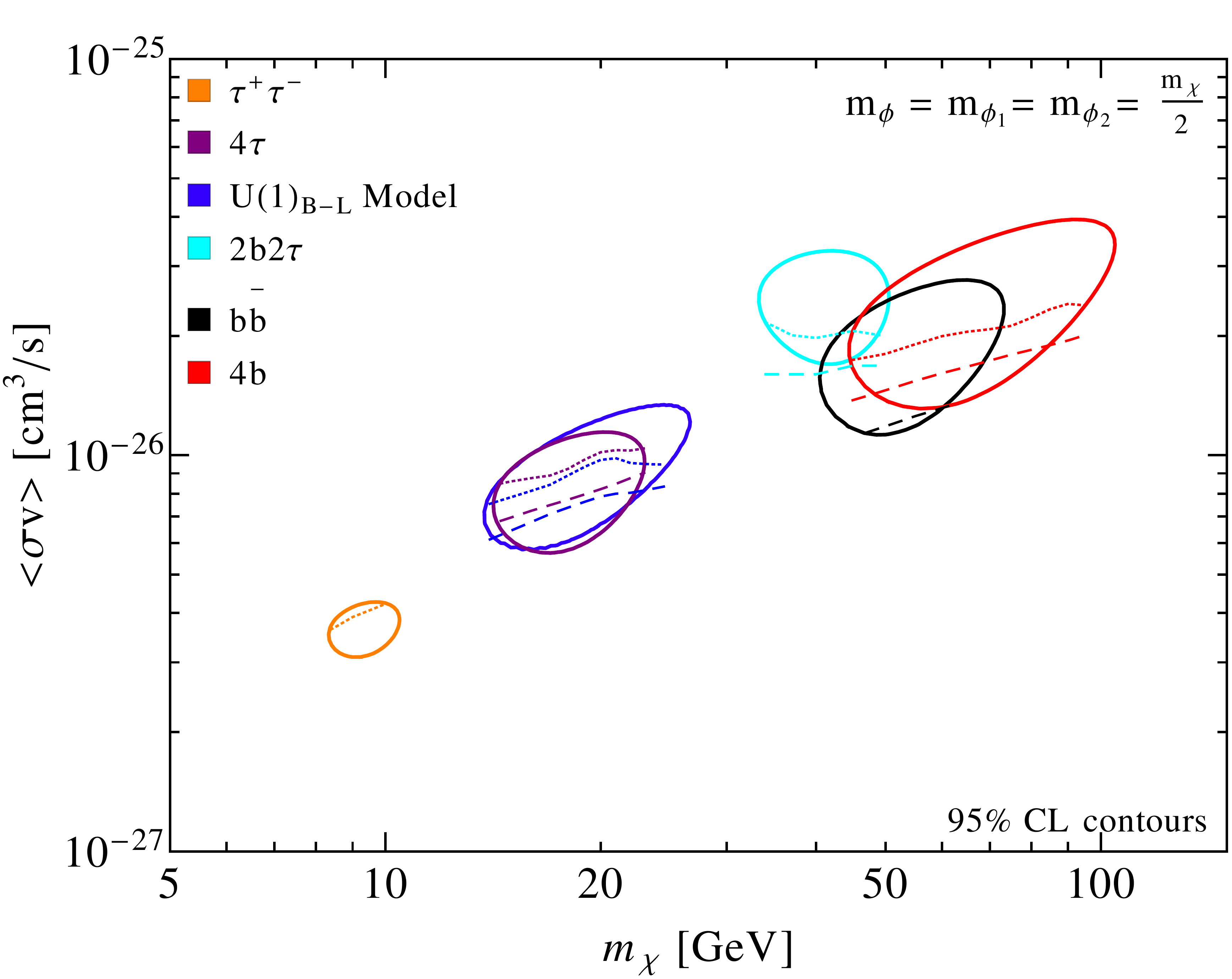}

\caption{Thermally averaged annihilation cross-sections (95 $\%$ CL contours) versus DM mass that fit the GCE. The dashed (dotted) lines are dSph constraints derived by scaling from $b\bar{b}$ ($\tau^+\tau^-$) limits provided by Fermi-LAT.  Regions below dashed (dotted) lines are still allowed, and $m_{\phi} = m_{\phi_1} = m_{\phi_2}$ is set to $m_{\chi}/2$. The contour for the $U(1)_{B-L}$ model, described in Section~\ref{Model}, are shown in dark blue. }
\label{dSphs-bound}
\end{figure}

In Figure~\ref{dSphs-bound} we show the annihilation cross-section and the DM mass for different channels that fit the GCE at 95$\%$ CL. As mentioned in the previous section the upper bounds on the annihilation cross-section of 4-body final states are derived by scaling from upper bounds of both $b \bar{b}$ and $\tau^+ \tau^-$ final states provided by Fermi-LAT. We find the bounds calculated from scaling to $b \bar{b}$ data are stronger than the same bounds computed from scaling to $\tau^+ \tau^-$. In both cases, we see that there are still regions of parameter space allowed by the dSph constraints. 

\begin{figure}[!t]
\centering
\includegraphics[scale=.3]{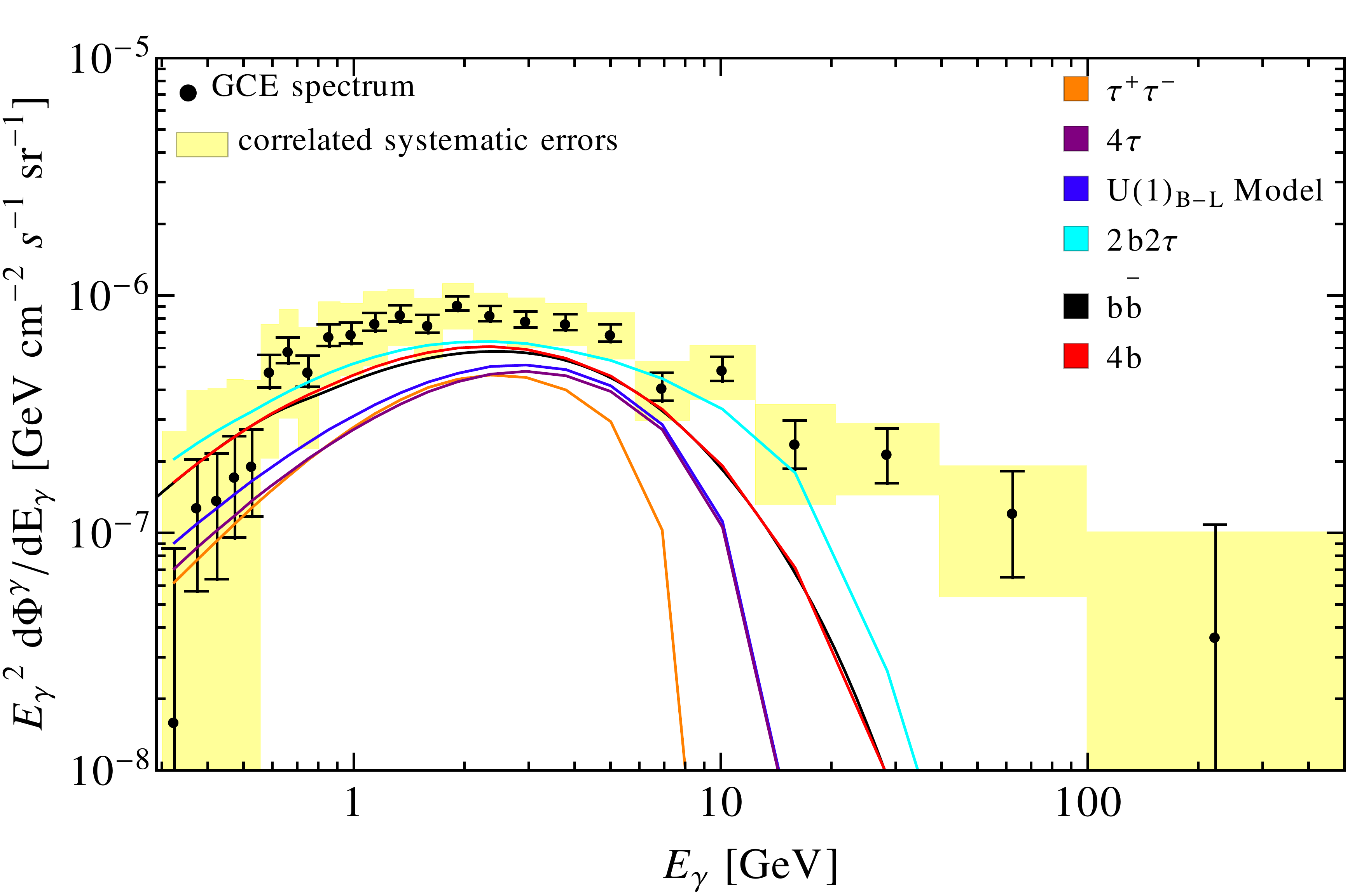}
\caption{
The photon spectra for the best-fit values of ($m_{\chi},\braket{\sigma v}$) for all channels shown in Table~\ref{tab1}.  
The GCE data together with statistical and systematic errors~\cite{Calore:2014xka} has been used and $m_{\phi} = m_{\phi_1} = m_{\phi_2}$ has been set to $m_{\chi}/2$ for these fits.}
\label{Spectra}
\end{figure}

The $\braket{\sigma v}$ values for the best-fit point for each channel are tabulated in Table~\ref{tab1}, along with the corresponding $\Delta \chi^2$ values representing a measure of the goodness-of-fit. The upper bound on $\braket{\sigma v}$ for these points from the dSph constraint is also shown in Table~\ref{tab1}, and the spectra for the best-fit points are shown in Figure~\ref{Spectra}. We note that the prompt photon spectra in Figure~\ref{Spectra} do not appear to be a good fit to the CCW data because only diagonal elements of the covariance matrix, $\Sigma_{ij}$, are depicted. The covariance matrix contains off-diagonal elements which are comparable to the diagonal elements because of the strong correlation of the systematic errors across energy bins. When the full covariance matrix is taken into account in the $\Delta \chi^2$ computation, the result is a more reasonable measure of goodness-of-fit, as shown in Table~\ref{tab1}. 

\begin{table}[b] 
\begin{center}
\begin{tabular}{| c| c|c c c| c c|} 

\hline
\multirow{3}{*}{Channel} & \multirow{3}{*}{$m_{\chi}$} & \multicolumn{3}{c|}{Best Fit} &  \multicolumn{2}{c|}{dSphs Allowed}\\
\cline{3-7}        

\rule{0pt}{3ex}        & & & & & $b \bar{b}$ & $\tau^+ \tau^-$ \\               

\cline{6-7}
        &            &   $\braket{\sigma v}$ & & $\Delta \chi^2_{min}$ & \multicolumn{2}{c|}{$\braket{\sigma v}_{max}$} \\
        &   (GeV)    &     ($10^{-26}$ cm$^3$s$^{-1}$) & & & \multicolumn{2}{c|}{($10^{-26}$cm$^3$s$^{-1}$)}  \\

\hline        

$\tau^+ \tau^-$ &  9 &  0.36 & & 33.4 & --   & 0.39 \\
$4 \tau$        & 19 &  0.90 & & 28.2 & 0.78 & 0.95 \\
$U(1)_{B-L}$    & 19 &  0.97 & & 27.5 & 0.75 & 0.91 \\
$ 2b \, 2 \tau$ & 41 &  2.43 & & 26.7 & 1.64 & 2.01\\
$b \bar{b}$     & 50 &  1.80 & & 25.2 & 1.18 &  --  \\
$  4 b$         & 65 &  2.45 & & 23.1 & 1.64 & 1.99   \\

\hline
\end{tabular}
\end{center} 
\caption{Best-fit results of spectral fits to the Fermi Galactic center excess in different channels, together with 95$\%$ CL limits and the upper bound on $\braket{\sigma v}$, coming from dSphs, for the corresponding point. The upper bounds on the annihilation cross-section of 4-body final states are derived by scaling from upper bounds of both $b \bar{b}$ and $\tau^+ \tau^-$ final states provided by Fermi-LAT and shown under columns named $b \bar{b}$ and $\tau^+ \tau^-$ respectively. $m_{\phi} = m_{\phi_1} = m_{\phi_2} = m_{\chi}/2$ is assumed for these points. The results for the $U(1)_{B-L}$ model are also shown in the Table.}
\label{tab1}
\end{table}

For the analysis in Figure~\ref{dSphs-bound} and Table~\ref{tab1}, the mass of the scalar mediators $m_{\phi} = m_{\phi_1} = m_{\phi_2}$ are set to $m_{\chi}/2$. To investigate the impact of the mass of these scalar mediators, we have fit to the GCE and evaluated the corresponding $\Delta\chi^2$, with different $m_{\phi}$ values for our best-fit points in all 4-body channels. We show these results in Table~\ref{tab2} for the $4b$ and $4 \tau$ channels. From Table~\ref{tab2} it is evident that the best-fit to the GCE is obtained for $m_{\phi} \sim m_{\chi}/2$. We verified that the conclusions are similar for the $2b \, 2\tau$ channel as well.  

Our results agree with previous studies in that light mediators are favored by the GCE~\cite{Hooper-4body,4body}. However, we favor a lower boost-factor, which is defined as $\gamma_{boost} \simeq 4m^2_{\chi} /( 4 m_{\phi} m_{\chi})$. For instance, Ref.~\cite{Hooper-4body} shows that the GCE  prefers $m_{\phi} \sim 2 m_b$ (with $\gamma_{boost} \sim 7$) or $m_{\phi} \sim m_{\chi}$ using the data of Ref.~\cite{Daylan:2014rsa}. For comparison our analysis shows $\gamma_{boost} \sim 2$ is preferred by the GCE. 

\begin{figure}[!t]
\centering
\includegraphics[scale=.2]{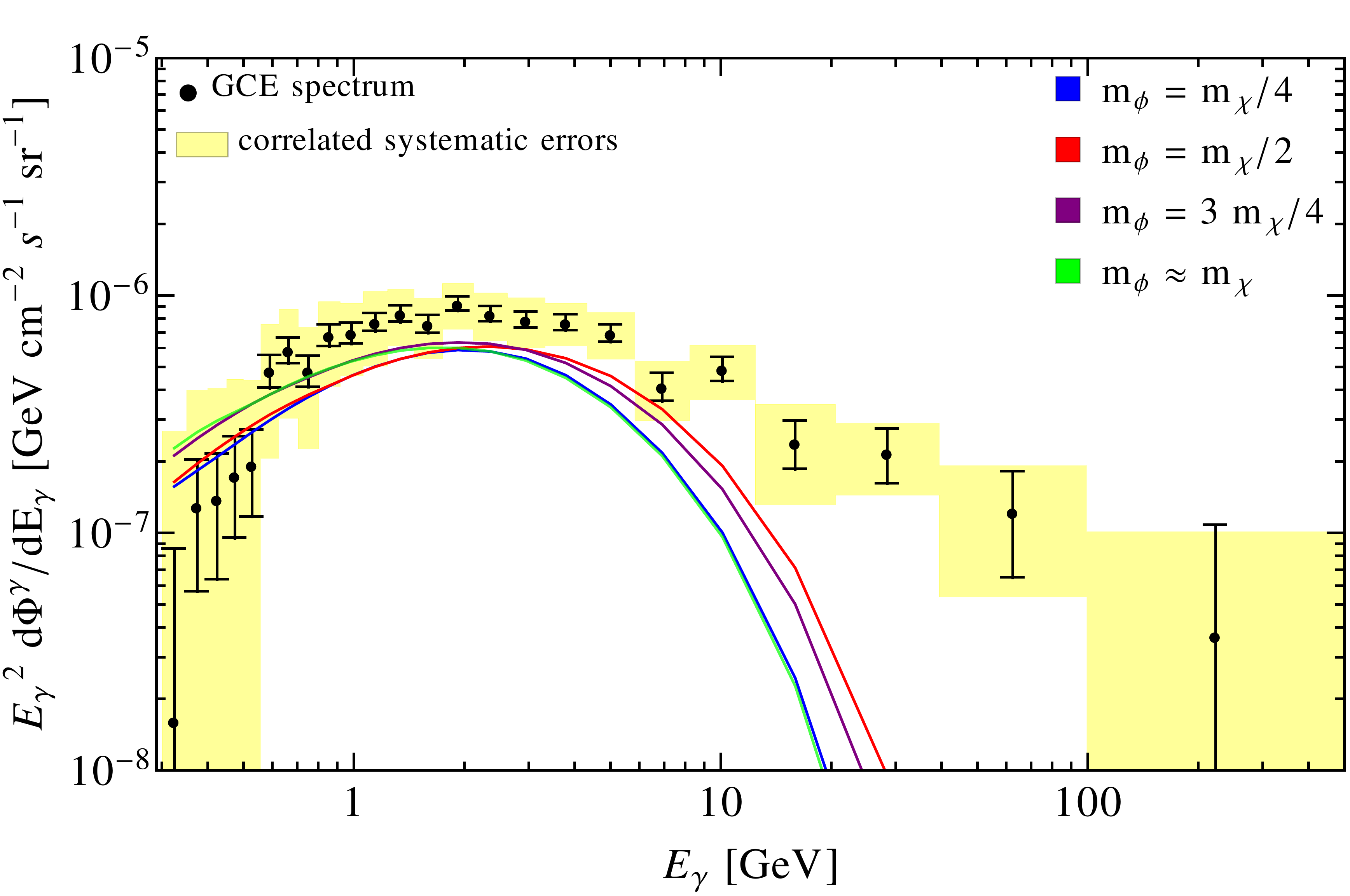}
\includegraphics[scale=.2]{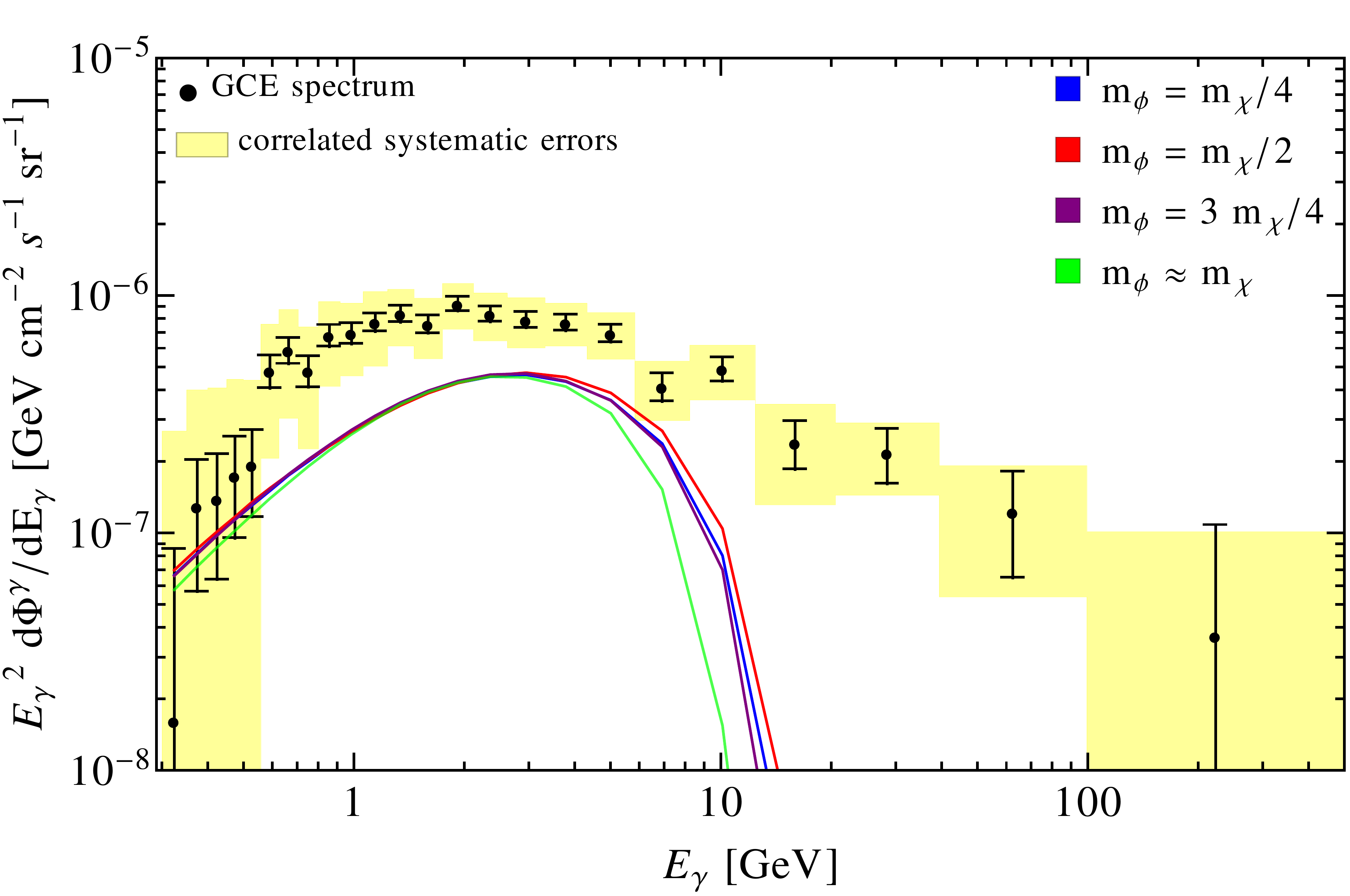}
\caption{The photon spectra in $4b$ [\textit{Left}] and $4 \tau$ [\textit{Right}] final states for different values of mediator mass $m_{\phi}$. $m_{\chi}$ has been set to its best-fit value of 65 GeV (19 GeV) for $4b$ ($4 \tau$) channel and best-fit $\braket{\sigma v}$ for each value of $m_{\phi}$ has been used in this plot.}
\label{Phi-spectra}
\end{figure}

We find that by including the correlated systematics of CCW, the GCE is better fit by a relatively broad spectrum for $m_{\phi} \sim m_{\chi}/2$. 
Different masses of $\phi$ broaden out the spectrum, as is illustrated in Fig.~\ref{Phi-spectra} for the best-fit points ($m_{\chi} = 65$ GeV and 19 GeV) in $4b$ and $4\tau$ final states respectively. 
Evidently the output of the CCW data is best fit by the broadest spectrum arising for $m_{\phi} = m_{\chi}/2$. For comparison the narrower spectra of $m_{\phi} = m_{\chi}/4$ and $m_{\phi} \approx m_{\chi}$ do not provide as good of a fit to the data. For a more detailed comparison of these two data sets, together with Fermi-LAT's analysis of the GCE~\cite{Fermi-GCE}, we refer to Refs.~\cite{Fermi-pass8,Agrawal:2014oha,Cline:2015qha}.

\begin{table}[!htp] 
\begin{center}
\begin{tabular}{| c|c c c| c c c|} 

\hline
\multirow{3}{*}{$m_{\phi}$} & \multicolumn{6}{c|}{Best Fit} \\
 \cline{2-7}     
 
  & \multicolumn{3}{c|}{$4b$} & \multicolumn{3}{c|}{$4 \tau$}\\
\cline{2-7}        


 \rule{0pt}{3ex}      &   $\braket{\sigma v}$ & & $\Delta \chi^2_{min}$ & $\braket{\sigma v}$ & & $\Delta \chi^2_{min}$\\
    (GeV)   &     ($10^{-26}$ cm$^3$s$^{-1}$) & & &     ($10^{-26}$ cm$^3$s$^{-1}$) & & \\
\hline        

$m_{\chi}/4$ & 2.14 & & 27.7 & 0.86 & & 29.0 \\
$m_{\chi}/2$ & 2.45 & & 23.1 & 0.90 & & 28.2 \\
$3 m_{\chi}/4$ & 2.58 & & 26.3 & 0.86 & & 28.9 \\
$\approx m_{\chi}$ & 2.39 & & 33.7 & 0.78 & & 32\\

\hline
\end{tabular}
\end{center} 
\caption{The dependence of goodness-of-fit on the mass of the scalar mediator, $\phi$, for the best-fit point ($m_{\chi}=65$ GeV and 19 GeV) in $4b$ and $4 \tau$ channels respectively. 
}
\label{tab2}
\end{table}

Let us now consider the results from Table~\ref{tab1} and Fig.~\ref{dSphs-bound} in more detail. From Table~\ref{tab1} we  notice that the $\phi \phi \rightarrow 4b$ final state offers the best-fit to the CCW GCE data, with a best fit mass and cross section of $m_{\chi} = 65$ GeV and $\braket{\sigma v} = 2.45 \e{-26}$ cm$^3$/s, respectively. The $b\bar{b}$ final state also provides a good-fit to the data for $m_{\chi} = 50$ GeV and $\braket{\sigma v} = 1.80 \e{-26}$ cm$^3$/s. The GCE can be explained by a wide range of DM mass for $\braket{\sigma v} \sim 1.27-4 \e{-26}$ cm$^3$/s ($4b$ channel) and $\braket{\sigma v} \sim 1.08-2.76 \e{-26}$ cm$^3$/s ($2b$ channel) respectively at 95$\%$ CL. The $4b$ channel allows the widest range of DM mass, 45-103 GeV, while $b\bar{b}$ fits for 43-73 GeV. The dSph constraint also allows a larger area in the $m_{\chi} - \braket{\sigma v}$ plane for the $4b$ final state as compared to the $b \bar{b}$ final state. However for both channels the best-fit $\braket{\sigma v}$ value from the GCE is disallowed by the dSph constraint. The $\phi_1 \phi_2 \rightarrow 2b 2\tau$ final state fits the data as well for $m_{\chi} \sim 37-50$ GeV with $\braket{\sigma v} \sim 1.53 - 3.3 \e{-26}$ cm$^3$/s at 95$\%$ CL, with a best-fit obtained for $m_{\chi} = 41$ GeV. The dSph constraint derived from $b\bar{b}$ rules out  the entire $m_{\chi} - \braket{\sigma v}$ plane in the parameter space, while the constraint from $\tau^+\tau^-$ allows a small fraction of it.  

In a similar manner, the $\phi \phi \rightarrow 4\tau$ final state allows for a wider range of DM mass (15-23 GeV), as opposed to a very narrow window (8.4-10.4 GeV) for the $\tau^+ \tau^-$ final state. There is a slightly larger range of $\braket{\sigma v}$ values preferred by the $4\tau$ channel ($0.56-1.19\e{-26}$ cm$^3$/s) relative to the $2 \tau$ channel ($0.31-0.43\e{-26}$ cm$^3$/s). However, while almost the entire parameter space is allowed for the $\tau^+ \tau^-$ final state for our choice of $\gamma$
and $\rho_0$ (previously observed by Ref.~\cite{Fermi-pass8}), an appreciable area of the $4 \tau$ final state is ruled out by the dSph constraint. 


To this point we have neglected two effects which may have an impact on our results. This first is the effect of Inverse Compton scattering (ICS). ICS modifies the observed photon spectrum, typically by $\sim 10 \%$ for final states involving $\tau$'s. It is less significant when considering final states with $b$-quarks. The impact of ICS on the GCE has been discussed in detail~\cite{Calore:2014nla}

\begin{figure}[!t]
\centering
\includegraphics[scale=.2]{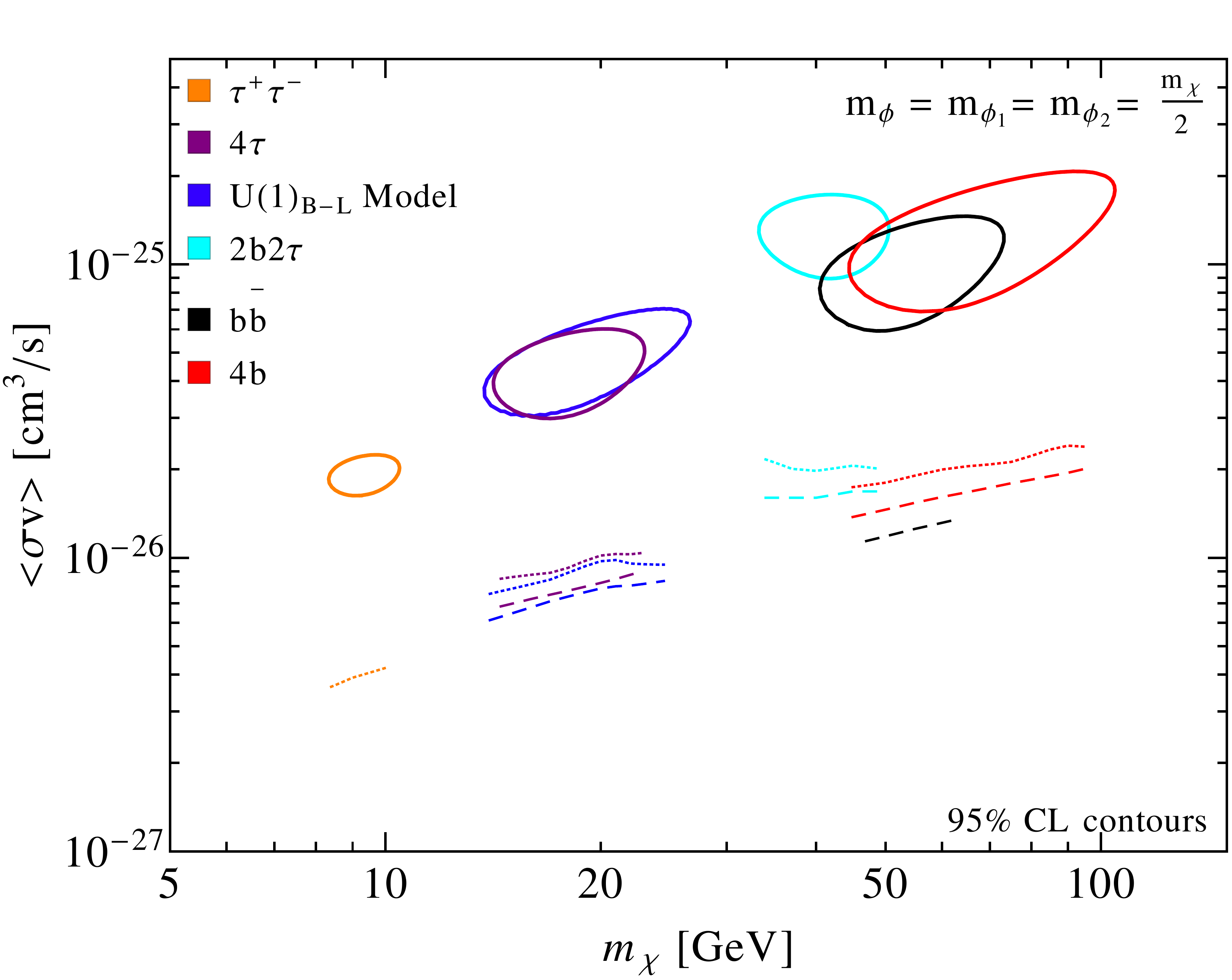}
\includegraphics[scale=.2]{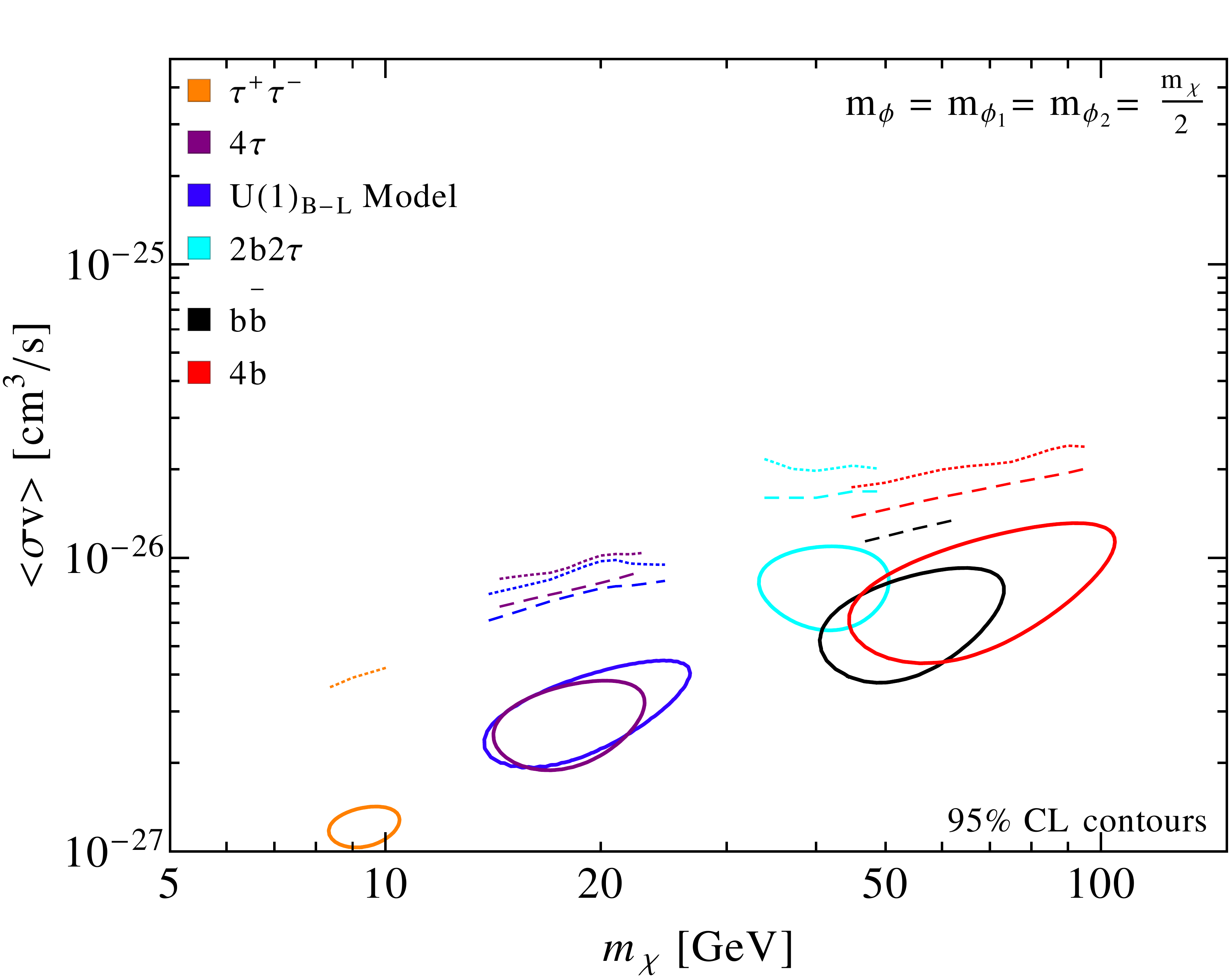}
\caption{Thermally averaged annihilation cross-sections (95 $\%$ CL contours) versus DM mass that fit the GCE, along with corresponding dSphs constraints if extreme values of J-factor are used. In the \textit{Left} panel $\bar{J} = 4 \times10^{22}$ GeV$^2$cm$^{-5}$sr$^{-1}$ ($\rho_\odot = 0.2$ GeV/cm$^3$, $\gamma=1.1$), and in the \textit{Right} panel  $6.07\times10^{23}$ GeV$^2$cm$^{-5}$sr$^{-1}$ ($\rho_\odot = 0.6$ GeV/cm$^3$, $\gamma=1.3$). Dashed and dotted lines representing dSph constraints have the same meaning as in Fig.~\ref{dSphs-bound}. Here we also take $m_{\phi} = m_{\phi_1} = m_{\phi_2}$ is set to $m_{\chi}/2$.}
\label{J-factor}
\end{figure}

\begin{figure}[!t]
\centering
\includegraphics[scale=.3]{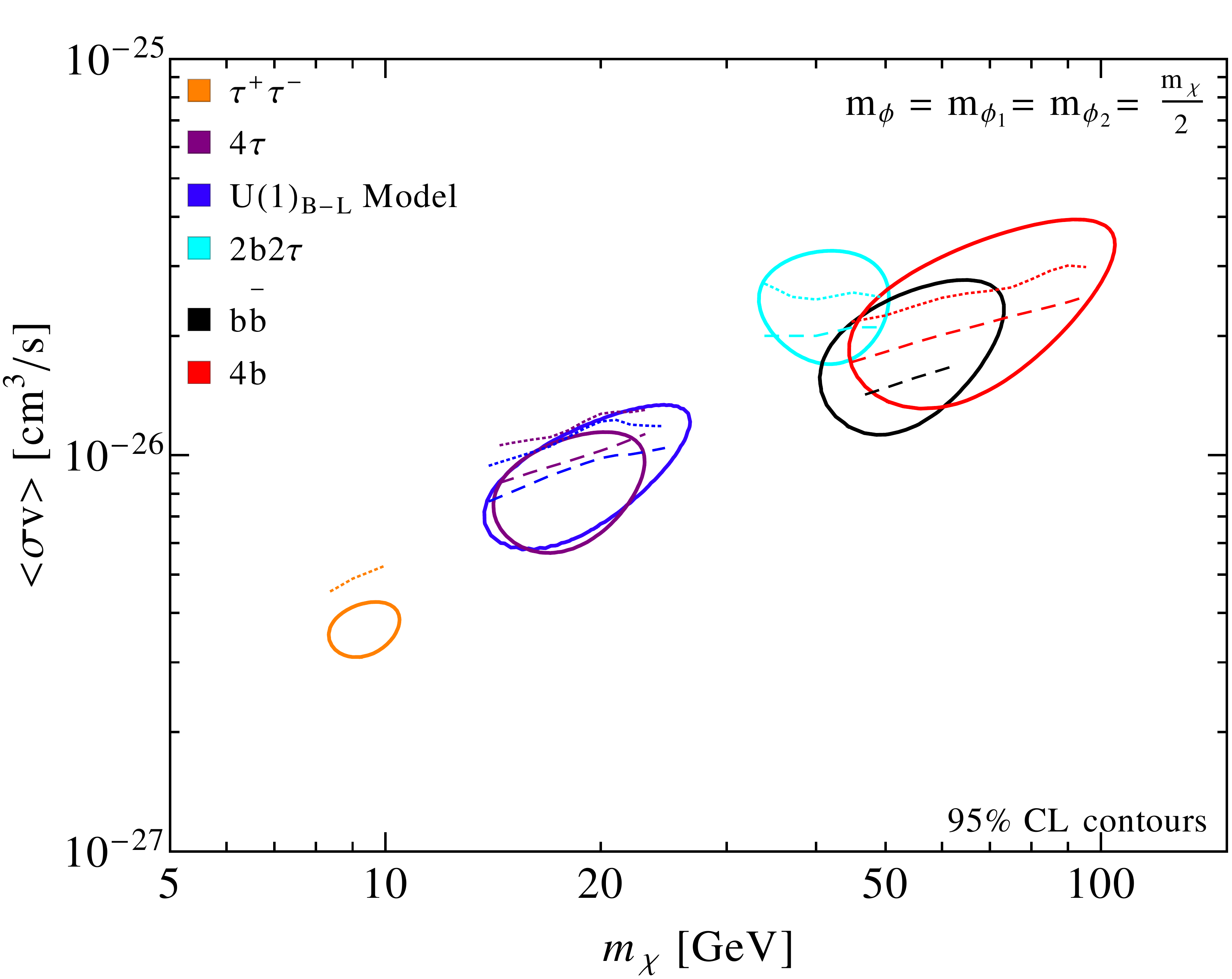}
\caption{The dSph constraints on the parameter space that fits GCE, if Burkert DM profile is used for dSphs. Dashed and dotted lines representing dSph constraints have the same meaning as in Fig.~\ref{dSphs-bound}. As above we take $m_{\phi} = m_{\phi_1} = m_{\phi_2}$ is set to $m_{\chi}/2$.}
\label{Burkert}
\end{figure}

In addition the results we have presented assume gNFW profile, which is shown to provide the best fit for the GCE~\cite{Daylan:2014rsa,Calore:2014xka}. There is of course a significant uncertainty in the average J-factor, $\bar{J}$, within the ROI that we consider, because of the uncertain DM distribution near the Galactic center. The impact of this uncertainty has been previously quantified, so that $\bar{J}$ anywhere from 0.19 to 3 times the canonical value that we use is allowed~\cite{Agrawal:2014oha}. The impact of this variation in J-factor on our fit to the GCE and the corresponding dSph constraint is shown in Figure~\ref{J-factor}. We conclude from Fig.~\ref{J-factor} that, with different choices for the DM profile, the dSph constraint can either exclude or allow the entire parameter space that fits GCE. Also assuming a Burkert model for the dSphs relaxes this bound by $\sim 25$\%~\cite{Ackermann:2015zua}. The effect of these relaxed bounds on our results are presented in Figure~\ref{Burkert}.

\section{\label{Model} $U(1)_{B-L}$ Model}

Now we move on to consider an example of an extended MSSM model where right-handed sneutinos is the DM candidate. This model provides a case of a cascade model to complement the model-independent approach highlighted above. The well motivated $B-L$ extension of the MSSM~\cite{mohapatra}  explains the neutrino masses and mixings since it has three right-handed neutrinos. The minimally extended model contains two new Higgs fields $H^{\prime}_1$ and $H^{\prime}_2$, the RH neutrinos $N$ together with their supersymmetric partners and a new gauge boson $Z^{\prime}$. The superpotential is $
W = W_{\rm MSSM} + W_{B-L} + y_D {\bf N}^c {\bf H_u} {\bf L} \,, $
where ${\bf L}$  and ${\bf H_u}$ are the superfields, that contain the Higgs field and provides mass to the left-handed leptons and up-type quarks respectively. The $W_{B-L}$ term consists of  ${\bf H^{\prime}_1},~{\bf H^{\prime}_2}$ and ${\bf N}^c$.  Charge assignments of the new Higgs fields determine the detailed form of $W_{B-L}$, e.g.,

\begin{center}
\begin{tabular}{|c||c|c|c|c|c|c|c|}\hline
{\rm Fields} & $Q$ & $Q^c$ & $L$ & $L^c$ &  $H^{\prime}_1$ & $H^{\prime}_2$ \\ \hline
$Q_{B-L}$ & 1/6 & -1/6 & -1/2 & 1/2 &  3/2 & -3/2 \\ \hline
\end{tabular}\end{center}
\vspace{0.2cm}

The scalar potential comprise of $F$-terms from the superpotential, and $D$-terms from the gauge symmetries. The $D$-term contribution from $U(1)_{B-L}$ is given by $
V_D \supset \frac{1}{2} D^2_{B-L}$, where
$
D_{B-L} = \frac{1}{2} g_{B-L} \left[Q_{1} ({\vert H^{\prime}_1 \vert}^2 - {\vert H^{\prime}_2 \vert}^2) + \frac{1}{2} {\vert \tilde N \vert}^2 + ... \right]. $
Here $g_{B-L}$ is the gauge coupling of $U(1)_{B-L}$, and $+Q_1$, $-Q_1$, $1/2$ are the $B-L$ charges of $H^{\prime}_1,~H^{\prime}_2,~{\tilde N}$ respectively (${\tilde N}$ is the sneutrino field). The $U(1)_{B-L}$ is broken by the vacuum expectation value (VEV) of $H^{\prime}_1$ and $H^{\prime}_2$, which we denote by $v^{\prime}_1$ and $v^{\prime}_2$ respectively. This results in a mass $m_{Z^\prime} =g_{B-L} Q_1 \sqrt{v^{\prime 2}_1 + v^{\prime 2}_2}$ for the $Z^{\prime}$ gauge boson. There are three physical Higgs fields $\phi,~\Phi$ (scalars) and ${\cal A}$ (a pseudo scalar). This scalar potential leads to the coupling between the right sneutrinos and the new Higgs particles.

The sneutrino,  ${\tilde N}$, is a natural candidate for DM in this model~\cite{dutta} (the lightest neutralino in the extended sector, which is a superposition of the two Higgsinos ${\widetilde H}^{\prime}_1$, ${\widetilde H}^{\prime}_2$ and  the $U(1)_{B-L}$ gaugino ${\widetilde Z}^{\prime}$, can also be a possible candidate~\cite{ADRS,khalil}.).   The dominant channel of the DM particle is ${\tilde N}^* {\tilde N} \rightarrow \phi \phi$ via the $s$-channel exchange of the $\phi,~\Phi$, the $t$,~$u$-channel exchange of the ${\tilde N}$, and the contact term $\vert {\tilde N} \vert^2 \phi^2$, where $\phi$ is the lightest scalar. The $s$-channel $Z^\prime$ exchange is subdominant because of the large $Z^\prime$ mass (as required by the experimental bound on $m_{Z^\prime}$).
There are also ${\tilde N}^* {\tilde N} \rightarrow \phi \Phi , ~ \phi {\cal A}, ~ \Phi \Phi,~{\cal A} {\cal A}$ annihilation processes, but they are kinematically suppressed and/or forbidden for the parameter space we are considering. The sneutrinos can also annihilate to RH neutrinos via $t$-channel neutralino exchange. Again for the parameter space that we consider the annihilation into $\phi \phi$ final states is dominant. Other fermion final states, through $s$-channel $Z^\prime$ exchange, have even smaller branching ratios. The annihilations to fermion-antifermion final states are $p$-wave suppressed.

\begin{figure}[!t]
\centering
\includegraphics[scale=.3]{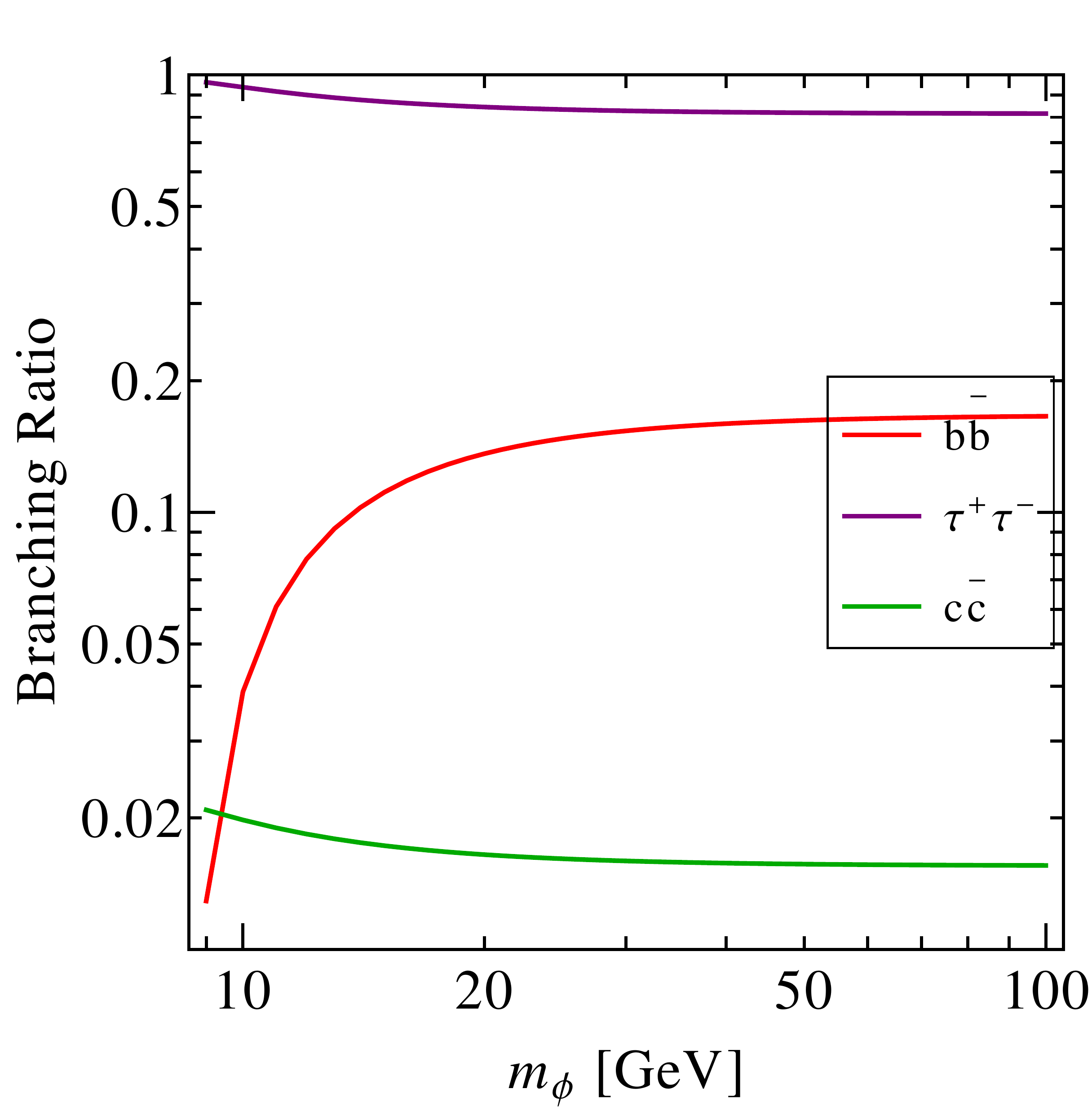}

\caption{Branching Ratio of $\phi$ in different fermion-antifermion pairs as a function of mass.} 
\label{BR}
\end{figure}

The $\phi$ subsequently decays into fermion-antifermion pairs via a one-loop diagram containing two $Z^{\prime}$ bosons. The decay rate is given by $
\Gamma(\phi \to f {\bar f}) = \frac{C_f}{2^7 \pi^5} \frac{g_{B-L}^6 Q^4_{f} Q^2_{\phi} m^5_{\phi} m^2_f}{m^6_{Z^{\prime}}} \left(1 - \frac{4 m^2_f}{m^2_{\phi}} \right)^{3/2}$,
where $Q_f$ and $Q_\phi$ are the $B-L$ charges of the final state fermion and the $\phi$ respectively, $m_f$ is the fermion mass, and $C_f$ denotes color factor ~\cite{dutta,ADRS}.  Evidently the leptonic BR is larger than that for quarks due to three times larger $B-L$ charge of leptons cmpared to  quarks. We should point out that $m_{\phi}$ is regulated by the VEVs of the new Higgs fields  and for $\tan \beta^{\prime} \approx 1$, i.e. when the VEVs are comparable,  it can be very small compared to $m_{Z^\prime}$. For $m_{\phi} > 2 \, m_b$ the dominant decay mode is $\phi \rightarrow \tau^{-} \tau^{+}$ , while the BR for the $\phi \rightarrow b {\bar b}$ mode is $\approx 7$ times smaller. The BR of $\phi$ as a function of $m_{\phi}$ is shown in Fig.~\ref{BR}, with $m_{Z'}=2.1$ TeV~\cite{mZp} and $g_{B-L}=0.4$. For the mass range of interest for the GCE study, the model is dominated by decay of $\phi$ to $\tau^+ \tau^-$ pair with  $\sim 80-90 \%$ BR.

Using reasonable values for the model parameters, i.e., $\tan \beta^\prime \approx 1$, $m_{Z^\prime} > 1.5$~TeV, $\mu^\prime = 0.5 - 2$~TeV ($\mu^{\prime}$ being the Higgs mixing parameter in the $B-L$ sector), soft masses for the Higgs fields $m_{H_{1,2}^\prime} = 200-600$~GeV, and soft gaugino mass $M_{\widetilde{Z}^\prime} \geq 500$~GeV, we find that the thermal relic abundance can be satisfied in this model with the DM mass, $m_{\tilde N}\sim 10-60$ GeV which we will use for our analysis. Since we consider ${\tilde N}^* {\tilde N} \rightarrow \phi \phi$, $m_{\phi}$ is smaller than the DM mass. We use $g_{B-L} \sim 0.3-0.40$, which is in concordance with unification of the gauge couplings~\cite{dutta}. The large $Z^\prime$ mass in this model also allows us to satisfy the direct detection~\cite{lux} and collider bounds~\cite{tev,carena,lhc}.

We are now in position to describe our results for the $U(1)_{B-L}$ model, which includes all 4-body channels discussed in the Section~\ref{Results}. The DM annihilation spectra arising from the model is dictated by the $\tilde{N}^* \tilde{N} \rightarrow \phi \phi \rightarrow 4 \tau$ process with $80-90 \%$ probability. Hence one may suspect that the set of ($m_{\tilde{N}}, \braket{\sigma v}$) values of the model that fit the GCE should be very similar to $4 \tau$ case and this fact is demonstrated in  Fig.~\ref{dSphs-bound} and Table~\ref{tab1}. Similar to the $4\tau$ case the best-fit to the GCE is obtained for $m_{\tilde{N}} = 19$ GeV. Although $\tilde{N}$ mass of 14.5-25 GeV fit the GCE excess with $\braket{\sigma v} \sim 0.58-1.42 \e{-26}$ cm$^3$/s, only a fraction of it is allowed by dSphs. 

Though we discuss the B-L model for the  DM annihilation to 4$b$, 4$\tau$, $2b\, 2\tau$ final states with specific BRs of $\phi\rightarrow b{\bar b}$ and $\phi\rightarrow \tau^+ \tau^-$ determined by the B-L charges, the analysis that we have presented in the previous two sections can be used for other models since we show our results for  generic BRs of the scalar state $\phi$ to $b\bar b$ and $\tau^+ \tau^-$ final states.

\section{\label{Conclusion}Conclusion}

In this paper we performed a model-independent fit to the GCE for DM particles annihilating to $4b$, $4 \tau$ and $2b \, 2\tau$ final states by means of cascade annihilation through a pair of BSM scalars $\phi$ (two scalars $\phi_1,\phi_2$ for  $2b \, 2\tau$ final state). We compared these results with standard $b \bar{b}$ and $\tau^+ \tau^-$ final states. We also presented a well motivated $U(1)_{B-L}$ model, where the lightest right-handed sneutrino ($\tilde{N}$) is the DM candidate, which provides a realistic scenario incorporating all 4-body final states mentioned above. The main result of this paper is the constraint imposed on the $m_{\chi} - \braket{\sigma v}$ plane for aforementioned 4-body channels by the reprocessed Fermi-LAT {\tt Pass-8} data on dwarf spheroidal galaxies. 

We found a wide range of DM masses that fit the GCE in 4-body final states with distinct range of annihilation cross-sections characteristic of the final state. However a considerable area of the $m_{\chi} - \braket{\sigma v}$ plane is disallowed by the dSph constraint, strongly constraining the DM interpretation of the GCE. The scalar masses have limited impact on the analysis but $m_{\phi} \sim m_{\chi}/2$ provides the best fit to the spectra. The impact of ICS is also negligible for the final states under consideration.

The $4b$ channel provides the best-fit for $m_{\chi} \sim 45-103$ GeV and $\braket{\sigma v} \sim 1.27-4 \e{-26}$ cm$^3$/s at 95$\%$ CL with the upper-half of the parameter space ruled out by dSphs. The $2b \, 2\tau$ channel fits the excess for $m_{\chi} \sim 37-50$ GeV and $\braket{\sigma v} \sim 1.53-3.3 \e{-26}$ cm$^3$/s. Compared to the $4b$ final state, the dSphs are found to be considerably more constraining for  $2b \, 2\tau$ and $b \bar{b}$ channels. On the other hand they are a less stringent constraint for the $4 \tau$ final state. Out of the 95$\%$ CL fit of $m_{\chi} \sim 15-23$ GeV and $\braket{\sigma v} \sim 0.56-1.19 \e{-26}$ cm$^3$/s, a large area in the $m_{\chi} - \braket{\sigma v}$ plane remains available if bounds are derived by scaling from the $\tau^+ \tau^-$ channel. However the $\tau^+ \tau^-$ channel remains unconstrained. The $U(1)_{B-L}$ model mostly follows the $4 \tau$ channel for $m_{\tilde{N}}\sim 14.5-25$ GeV, which fits the GCE.

\bigskip
{\bf Acknowledgement}

We thank Tracy Slatyer for helpful discussions. This work  is supported by DOE Grant DE-FG02-13ER42020,  NSF Grant No. PHY-1522717 and support from the Mitchell Institute. Y.G. thanks the Mitchell Institute for Fundamental Physics and Astronomy for support.


\begin{thebibliography}{99}

%

\bibitem{Hooperon} 
  L.~Goodenough and D.~Hooper,
  arXiv:0910.2998 [hep-ph].
  D.~Hooper and L.~Goodenough,
  Phys.\ Lett.\ B {\bf 697}, 412 (2011)
  [arXiv:1010.2752 [hep-ph]].
  
  
\bibitem{GCE-others} 
  A.~Boyarsky, D.~Malyshev and O.~Ruchayskiy,
  Phys.\ Lett.\ B {\bf 705}, 165 (2011)
  [arXiv:1012.5839 [hep-ph]];
  D.~Hooper and T.~Linden,
  Phys.\ Rev.\ D {\bf 84}, 123005 (2011)
  [arXiv:1110.0006 [astro-ph.HE]];
  K.~N.~Abazajian and M.~Kaplinghat,
  Phys.\ Rev.\ D {\bf 86}, 083511 (2012)
  [Erratum-ibid.\ D {\bf 87}, 129902 (2013)]
  [arXiv:1207.6047 [astro-ph.HE]];
  B.~Zhou, Y.~F.~Liang, X.~Huang, X.~Li, Y.~Z.~Fan, L.~Feng and J.~Chang,
  arXiv:1406.6948 [astro-ph.HE].
    
\bibitem{Daylan:2014rsa} 
  T.~Daylan, D.~P.~Finkbeiner, D.~Hooper, T.~Linden, S.~K.~N.~Portillo, N.~L.~Rodd and T.~R.~Slatyer,
  arXiv:1402.6703 [astro-ph.HE].

\bibitem{Calore:2014xka} 
  F.~Calore, I.~Cholis and C.~Weniger,
  arXiv:1409.0042 [astro-ph.CO].
  
  C.~Gordon and O.~Macias,
  Phys.\ Rev.\ D {\bf 88}, no. 8, 083521 (2013)
  [Erratum-ibid.\ D {\bf 89}, no. 4, 049901 (2014)]
  [arXiv:1306.5725 [astro-ph.HE]].
    
\bibitem{Pulsar} 
  K.~N.~Abazajian,
  JCAP {\bf 1103}, 010 (2011)
  [arXiv:1011.4275 [astro-ph.HE]];
  C.~Gordon and O.~Macias,
  Phys.\ Rev.\ D {\bf 88}, no. 8, 083521 (2013)
  [Erratum-ibid.\ D {\bf 89}, no. 4, 049901 (2014)]
  [arXiv:1306.5725 [astro-ph.HE]].
  K.~N.~Abazajian, N.~Canac, S.~Horiuchi and M.~Kaplinghat,
  Phys.\ Rev.\ D {\bf 90}, no. 2, 023526 (2014)
  [arXiv:1402.4090 [astro-ph.HE]];
  Q.~Yuan and B.~Zhang,
  JHEAp {\bf 3-4}, 1 (2014)
  [arXiv:1404.2318 [astro-ph.HE]];
  T.~D.~Brandt and B.~Kocsis,
  arXiv:1507.05616 [astro-ph.HE].



  
\bibitem{O'Leary:2015gfa} 
  R.~M.~O'Leary, M.~D.~Kistler, M.~Kerr and J.~Dexter,
  arXiv:1504.02477 [astro-ph.HE].

  
\bibitem{Bartels:2015aea} 
  R.~Bartels, S.~Krishnamurthy and C.~Weniger,
  arXiv:1506.05104 [astro-ph.HE].
  
\bibitem{Lee:2015fea} 
  S.~K.~Lee, M.~Lisanti, B.~R.~Safdi, T.~R.~Slatyer and W.~Xue,
  arXiv:1506.05124 [astro-ph.HE].


\bibitem{Carlson:2014cwa} 
  E.~Carlson and S.~Profumo,
  Phys.\ Rev.\ D {\bf 90}, no. 2, 023015 (2014)
  [arXiv:1405.7685 [astro-ph.HE]].
  
\bibitem{Cholis:2015dea} 
  I.~Cholis, C.~Evoli, F.~Calore, T.~Linden, C.~Weniger and D.~Hooper,
  arXiv:1506.05119 [astro-ph.HE].

\bibitem{Petrovic:2014uda} 
  J.~Petrovic, P.~D.~Serpico and G.~Zaharijas,
  JCAP {\bf 1410}, no. 10, 052 (2014)
  [arXiv:1405.7928 [astro-ph.HE]].

\bibitem{Gaggero:2015nsa} 
  D.~Gaggero, M.~Taoso, A.~Urbano, M.~Valli and P.~Ullio,
  arXiv:1507.06129 [astro-ph.HE].
 
 
\bibitem{Calore:2014nla} 
  F.~Calore, I.~Cholis, C.~McCabe and C.~Weniger,
  Phys.\ Rev.\ D {\bf 91}, no. 6, 063003 (2015)
  [arXiv:1411.4647 [hep-ph]].

\bibitem{Anti-pulsar} 
  I.~Cholis, D.~Hooper and T.~Linden,
  arXiv:1407.5625 [astro-ph.HE].
  F.~Calore, M.~Di Mauro and F.~Donato,
  arXiv:1412.4997 [astro-ph.HE].
    

  


\bibitem{Agrawal:2014oha} 
  P.~Agrawal, B.~Batell, P.~J.~Fox and R.~Harnik,
  arXiv:1411.2592 [hep-ph].

\bibitem{Cerdeno:2015ega} 
  D.~G.~Cerdeno, M.~Peiro and S.~Robles,
  arXiv:1501.01296 [hep-ph].
  

\bibitem{Hooper-4body} 
  A.~Berlin, P.~Gratia, D.~Hooper and S.~D.~McDermott,
  Phys.\ Rev.\ D {\bf 90}, no. 1, 015032 (2014)
  [arXiv:1405.5204 [hep-ph]].

\bibitem{4body} 
  P.~Ko, W.~I.~Park and Y.~Tang,
  JCAP {\bf 1409}, 013 (2014)
  [arXiv:1404.5257 [hep-ph]];
  C.~Boehm, M.~J.~Dolan and C.~McCabe,
  Phys.\ Rev.\ D {\bf 90}, no. 2, 023531 (2014)
  [arXiv:1404.4977 [hep-ph]];
  M.~Abdullah, A.~DiFranzo, A.~Rajaraman, T.~M.~P.~Tait, P.~Tanedo and A.~M.~Wijangco,
  Phys.\ Rev.\ D {\bf 90}, no. 3, 035004 (2014)
  [arXiv:1404.6528 [hep-ph]];
  A.~Martin, J.~Shelton and J.~Unwin,
  Phys.\ Rev.\ D {\bf 90}, no. 10, 103513 (2014)
  [arXiv:1405.0272 [hep-ph]].
  

\bibitem{Cline:2015qha} 
  J.~M.~Cline, G.~Dupuis, Z.~Liu and W.~Xue,
  arXiv:1503.08213 [hep-ph].

\bibitem{GCE-DM} 
  L.~A.~Anchordoqui and B.~J.~Vlcek,
  Phys.\ Rev.\ D {\bf 88}, 043513 (2013)
  [arXiv:1305.4625 [hep-ph]];
  W.~C.~Huang, A.~Urbano and W.~Xue,
  JCAP {\bf 1404}, 020 (2014)
  [arXiv:1310.7609 [hep-ph]];
  A.~Hektor and L.~Marzola,
  Phys.\ Rev.\ D {\bf 90}, no. 5, 053007 (2014)
  [arXiv:1403.3401 [hep-ph]];
  A.~Alves, S.~Profumo, F.~S.~Queiroz and W.~Shepherd,
  Phys.\ Rev.\ D {\bf 90}, no. 11, 115003 (2014)
  [arXiv:1403.5027 [hep-ph]];
  A.~Berlin, D.~Hooper and S.~D.~McDermott,
  Phys.\ Rev.\ D {\bf 89}, no. 11, 115022 (2014)
  [arXiv:1404.0022 [hep-ph]];
  P.~Agrawal, B.~Batell, D.~Hooper and T.~Lin,
  Phys.\ Rev.\ D {\bf 90}, no. 6, 063512 (2014)
  [arXiv:1404.1373 [hep-ph]];
  E.~Izaguirre, G.~Krnjaic and B.~Shuve,
  Phys.\ Rev.\ D {\bf 90}, no. 5, 055002 (2014)
  [arXiv:1404.2018 [hep-ph]];
  D.~G.~Cerdeño, M.~Peiró and S.~Robles,
  JCAP {\bf 1408}, 005 (2014)
  [arXiv:1404.2572 [hep-ph]];
  S.~Ipek, D.~McKeen and A.~E.~Nelson,
  Phys.\ Rev.\ D {\bf 90}, no. 5, 055021 (2014)
  [arXiv:1404.3716 [hep-ph]];
  D.~K.~Ghosh, S.~Mondal and I.~Saha,
  JCAP {\bf 1502}, no. 02, 035 (2015)
  [arXiv:1405.0206 [hep-ph]];
  L.~Wang and X.~F.~Han,
  Phys.\ Lett.\ B {\bf 739}, 416 (2014)
  [arXiv:1406.3598 [hep-ph]];
  B.~D.~Fields, S.~L.~Shapiro and J.~Shelton,
  Phys.\ Rev.\ Lett.\  {\bf 113}, 151302 (2014)
  [arXiv:1406.4856 [astro-ph.HE]];
  C.~Arina, E.~Del Nobile and P.~Panci,
  Phys.\ Rev.\ Lett.\  {\bf 114}, 011301 (2015)
  [arXiv:1406.5542 [hep-ph]];
  C.~Cheung, M.~Papucci, D.~Sanford, N.~R.~Shah and K.~M.~Zurek,
  Phys.\ Rev.\ D {\bf 90}, no. 7, 075011 (2014)
  [arXiv:1406.6372 [hep-ph]];
  J.~Huang, T.~Liu, L.~T.~Wang and F.~Yu,
  Phys.\ Rev.\ D {\bf 90}, no. 11, 115006 (2014)
  [arXiv:1407.0038 [hep-ph]];
  P.~Ko and Y.~Tang,
  JCAP {\bf 1501}, 023 (2015)
  [arXiv:1407.5492 [hep-ph]];
  S.~Baek, P.~Ko and W.~I.~Park,
  arXiv:1407.6588 [hep-ph];
  N.~Okada and O.~Seto,
  Phys.\ Rev.\ D {\bf 90}, no. 8, 083523 (2014)
  [arXiv:1408.2583 [hep-ph]];
  K.~Ghorbani,
  JCAP {\bf 1501}, 015 (2015)
  [arXiv:1408.4929 [hep-ph]];
  N.~F.~Bell, S.~Horiuchi and I.~M.~Shoemaker,
  Phys.\ Rev.\ D {\bf 91}, no. 2, 023505 (2015)
  [arXiv:1408.5142 [hep-ph]];
  A.~D.~Banik and D.~Majumdar,
  Phys.\ Lett.\ B {\bf 743}, 420 (2015)
  [arXiv:1408.5795 [hep-ph]];
  D.~Borah and A.~Dasgupta,
  Phys.\ Lett.\ B {\bf 741}, 103 (2015)
  [arXiv:1409.1406 [hep-ph]];
  J.~H.~Yu,
  Phys.\ Rev.\ D {\bf 90}, no. 9, 095010 (2014)
  [arXiv:1409.3227 [hep-ph]];
  J.~Guo, J.~Li, T.~Li and A.~G.~Williams,
  arXiv:1409.7864 [hep-ph];
  J.~Cao, L.~Shang, P.~Wu, J.~M.~Yang and Y.~Zhang,
  Phys.\ Rev.\ D {\bf 91}, no. 5, 055005 (2015)
  [arXiv:1410.3239 [hep-ph]];
  M.~Heikinheimo and C.~Spethmann,
  JHEP {\bf 1412}, 084 (2014)
  [arXiv:1410.4842 [hep-ph]];
  K.~Cheung, W.~C.~Huang and Y.~L.~S.~Tsai,
  arXiv:1411.2619 [hep-ph];
  A.~Berlin, A.~DiFranzo and D.~Hooper,
  arXiv:1501.03496 [hep-ph];
  C.~H.~Chen and T.~Nomura,
  arXiv:1501.07413 [hep-ph];
  K.~P.~Modak and D.~Majumdar,
  arXiv:1502.05682 [hep-ph];
  A.~Achterberg, S.~Caron, L.~Hendriks, R.~Ruiz de Austri and C.~Weniger,
  arXiv:1502.05703 [hep-ph];
  A.~Berlin, S.~Gori, T.~Lin and L.~T.~Wang,
  arXiv:1502.06000 [hep-ph];
  T.~Gherghetta, B.~von Harling, A.~D.~Medina, M.~A.~Schmidt and T.~Trott,
  arXiv:1502.07173 [hep-ph];
  J.~Guo, Z.~Kang, P.~Ko and Y.~Orikasa,
  Phys.\ Rev.\ D {\bf 91}, no. 11, 115017 (2015)
  [arXiv:1502.00508 [hep-ph]];
  A.~Rajaraman, J.~Smolinsky and P.~Tanedo,
  arXiv:1503.05919 [hep-ph];
  E.~C.~F.~S.~Fortes, V.~Pleitez and F.~W.~Stecker,
  arXiv:1503.08220 [hep-ph];
  P.~Ko and Y.~Tang,
  arXiv:1504.03908 [hep-ph];
  J.~Kim, J.~C.~Park and S.~C.~Park,
  arXiv:1505.04620 [hep-ph];
  C.~Balázs, T.~Li, C.~Savage and M.~White,
  arXiv:1505.06758 [hep-ph];
  J.~Cao, L.~Shang, P.~Wu, J.~M.~Yang and Y.~Zhang,
  arXiv:1506.06471 [hep-ph];
  A.~Butter, T.~Plehn, M.~Rauch, D.~Zerwas, S.~Henrot-Versillé and R.~Lafaye,
  arXiv:1507.02288 [hep-ph];
  D.~Kim and J.~C.~Park,
  arXiv:1507.07922 [hep-ph];
  N.~Fonseca, L.~Necib and J.~Thaler,
  arXiv:1507.08295 [hep-ph];
  M.~R.~Buckley and D.~Feld,
  arXiv:1508.00908 [hep-ph];
  K.~P.~Modak, D.~Majumdar and S.~Rakshit,
  JCAP {\bf 1503}, 011 (2015)
  [arXiv:1312.7488 [hep-ph]];
  A.~D.~Banik, D.~Majumdar and A.~Biswas,
  arXiv:1506.05665 [hep-ph];
  A.~Biswas,
  arXiv:1412.1663 [hep-ph];
  A.~Biswas, D.~Majumdar and P.~Roy,
  JHEP {\bf 1504}, 065 (2015)
  [arXiv:1501.02666 [hep-ph]];
  E.~Hardy, R.~Lasenby and J.~Unwin,
  JHEP {\bf 1407}, 049 (2014)
  [arXiv:1402.4500 [hep-ph]].








\bibitem{Strigari:2013iaa} 
  L.~E.~Strigari,
  Phys.\ Rept.\  {\bf 531}, 1 (2013)
  [arXiv:1211.7090 [astro-ph.CO]].
  
\bibitem{Conrad:2015bsa} 
  J.~Conrad, J.~Cohen-Tanugi and L.~E.~Strigari,
  arXiv:1503.06348 [astro-ph.CO].

\bibitem{Ackermann:2015zua} 
  M.~Ackermann {\it et al.}  [Fermi-LAT Collaboration],
  arXiv:1503.02641 [astro-ph.HE].


\bibitem{Bergstrom:2012fi} 
  L.~Bergstrom,
  Annalen Phys.\  {\bf 524}, 479 (2012)
  [arXiv:1205.4882 [astro-ph.HE]].

\bibitem{gNFW} 
  J.~F.~Navarro, C.~S.~Frenk and S.~D.~M.~White,
  Astrophys.\ J.\  {\bf 462}, 563 (1996)
  [astro-ph/9508025];
   J.~F.~Navarro, C.~S.~Frenk and S.~D.~M.~White,
  Astrophys.\ J.\  {\bf 490}, 493 (1997)
  [astro-ph/9611107].





  
\bibitem{Bovy:2012tw} 
  J.~Bovy and S.~Tremaine,
  Astrophys.\ J.\  {\bf 756}, 89 (2012)
  [arXiv:1205.4033 [astro-ph.GA]].

\bibitem{Sjostrand:2014zea} 
  T.~Sjöstrand, S.~Ask, J.~R.~Christiansen, R.~Corke, N.~Desai, P.~Ilten, S.~Mrenna and S.~Prestel {\it et al.},
  arXiv:1410.3012 [hep-ph].


\bibitem{PPPC}
  M.~Cirelli, G.~Corcella, A.~Hektor, G.~Hutsi, M.~Kadastik, P.~Panci, M.~Raidal and F.~Sala {\it et al.},
  JCAP {\bf 1103}, 051 (2011)
  [Erratum-ibid.\  {\bf 1210}, E01 (2012)]
  [arXiv:1012.4515 [hep-ph], arXiv:1012.4515 [hep-ph]];
   P.~Ciafaloni, D.~Comelli, A.~Riotto, F.~Sala, A.~Strumia and A.~Urbano,
  JCAP {\bf 1103}, 019 (2011)
  [arXiv:1009.0224 [hep-ph]].

\bibitem{CCW} 
  F.~Calore, I.~Cholis and C.~Weniger,\\
\url{https://staff.fnwi.uva.nl/c.weniger/pages/material/}.

\bibitem{Ackermann:2013yva} 
  M.~Ackermann {\it et al.}  [Fermi-LAT Collaboration],
  Phys.\ Rev.\ D {\bf 89}, 042001 (2014)
  [arXiv:1310.0828 [astro-ph.HE]].

\bibitem{GeringerSameth:2011iw} 
  A.~Geringer-Sameth and S.~M.~Koushiappas,
  Phys.\ Rev.\ Lett.\  {\bf 107}, 241303 (2011)
  [arXiv:1108.2914 [astro-ph.CO]].

\bibitem{Ackermann:2011wa} 
  M.~Ackermann {\it et al.}  [Fermi-LAT Collaboration],
  Phys.\ Rev.\ Lett.\  {\bf 107}, 241302 (2011)
  [arXiv:1108.3546 [astro-ph.HE]].

\bibitem{Geringer-Sameth:2014qqa} 
  A.~Geringer-Sameth, S.~M.~Koushiappas and M.~G.~Walker,
  Phys.\ Rev.\ D {\bf 91}, 083535 (2015)
  [arXiv:1410.2242 [astro-ph.CO]].

\bibitem{Fermi-GCE}
S. Murgia, “Observation of the High Energy Gamma-ray Emission Towards the Galactic Center.”\\
\url{http://fermi.gsfc.nasa.gov/science/mtgs/symposia/2014/program/08_Murgia.pdf}


\bibitem{Fermi-pass8}
M. Wood, “New Fermi-LAT Results on the Search for Dark Matter Annihilation in Dwarf Spheroidal Galaxies.”\\
\url{https://indico.cern.ch/event/344116/material/slides/0.pdf}



\bibitem{mohapatra}
R.~N.~Mohapatra and R.~E.~Marshak,
Phys.\ Rev.\ Lett.\  {\bf 44}, 1316 (1980)
[Erratum-ibid.\  {\bf 44}, 1643 (1980)].
  



  
\bibitem{dutta} 
  R.~Allahverdi, B.~Dutta, K.~Richardson-McDaniel and Y.~Santoso,
  Phys.\ Lett.\ B {\bf 677}, 172 (2009)
  [arXiv:0902.3463 [hep-ph]].
  
\bibitem{ADRS}
R.~Allahverdi, B.~Dutta, K.~Richardson-McDaniel and Y.~Santoso,
  Phys.\ Rev.\ D {\bf 79}, 075005 (2009)
  [arXiv:0812.2196 [hep-ph]].

\bibitem{khalil}
S.~Khalil and H.~Okada,
  Phys.\ Rev.\ D {\bf 79}, 083510 (2009)
  [arXiv:0810.4573 [hep-ph]].

\bibitem{mZp} 
  A.~Alves, A.~Berlin, S.~Profumo and F.~S.~Queiroz,
  arXiv:1501.03490 [hep-ph].

\bibitem{lux} 
  D.~S.~Akerib {\it et al.}  [LUX Collaboration],
  Phys.\ Rev.\ Lett.\  {\bf 112}, 091303 (2014)
  [arXiv:1310.8214 [astro-ph.CO]].

\bibitem{tev}
T.~Aaltonen {\it et. al.},
  Phys.\ Rev.\ Lett.\  {\bf 99}, 171802 (2007).

\bibitem{carena}
M.~S.~Carena {\it et. al.},
  Phys.\ Rev.\  D {\bf 70}, 093009 (2004).

\bibitem{lhc}
  Y.~A.~Coutinho, E.~C.~F.~S.~Fortes and J.~C.~Montero,
  Phys.\ Rev.\ D {\bf 84}, 055004 (2011)
  [Phys.\ Rev.\ D {\bf 84}, 059901 (2011)]
  [arXiv:1102.4387 [hep-ph]].





  

  




\end{thebibliography}
\end{document}